\documentstyle[twoside,fleqn,espcrc2]{article}

\def \bc {\begin{center}}
\def \ec {\end{center}}

\def \bfr {\begin{flushright}}
\def \efr {\end{flushright}}

\def \ba {\begin{array}}
\def \ea {\end{array}}

\def \bea {\begin{eqnarray}}
\def \eea {\end{eqnarray}}

\def \be {\begin{equation}}
\def \ee {\end{equation}}

\def\nn{\nonumber}

\def\l[{\left[}
\def\r]{\right]}

\newcommand{\AmS}{{\protect\the\textfont2
  A\kern-.1667em\lower.5ex\hbox{M}\kern-.125emS}}

\title{Dynamical content of quantum diffeomorphisms in two-dimensional quantum 
gravity}

\author{ V. Aldaya
             and
	J.L. Jaramillo
\address{Instituto de Astrof\'{\i}sica 
de Andaluc\'{\i}a, CSIC, Apartado Postal 3004,
18080 Granada, Spain.}
        \thanks{Work partially
supported by the DGICYT.}}
       
\begin{document}

\begin{abstract}
A model for 2D-quantum gravity from the Virasoro symmetry is studied. The
notion of space-time naturally arises as a homogeneous space associated 
with the kinematical (non-dynamical) SL(2,R) symmetry in the kernel of the 
Lie-algebra central extension for the critical values of the conformal 
anomaly. The rest of the generators in the group, $L_{\mid n\mid\geq 2}$, 
mix space-times with different constant curvature.
Only in the classical limit all space-times can be identified, defining a 
unique Minkowski space-time, and the operators $L_{\mid n\mid\geq 2}$ gauged 
away. This process entails a restriction to SL(2,R) subrepresentations, 
which creates a non-trivial two-dimensional symplectic classical phase space. 
The present model thus suggests that the role of general covariance in 
quantum gravity is different from that played in the classical limit. 
\end{abstract}

\maketitle

\section{Introduction}
Our main {\it goal} is the construction of a space-time notion just from 
symmetry  principles. Therefore, we consider space-time as a derived, or 
secondary, object in our theory.
Assuming that the nature of the world is a quantum one, we want to {\it start}
from the very beginning with a quantum theory in which, perhaps, there is not
an explicit notion of space-time. We understand by such a quantum theory a 
unitary and irreducible representation of the algebra of physical operators 
and the space-time structure itself, in case such an structure makes any
sense, should be built out of this quantum theory.

In order to fill the gap between the starting point (quantum theory) and the 
objective (space-time construction), we present a very simple model that we 
interpret as a quantum gravity (see \cite{QG0899}), in fact a two-dimensional 
one, even though the link with standard two-dimensional quantum gravity is not
completely obvious.

The technical tool we are going to use is a Group Approach to Quantization 
(GAQ) (see \cite{Aldaya} and references therein), which starts from a given 
group of symmetries and constructs a quantum theory in the sense mentioned 
above. Among the elements of this approach, one is specially important for
this work as it will play the role of physical guide in our search of
space-time. This element is group (pseudo-)cohomology, which is a mathematical 
concept related to central extensions of the considered group. Its 
importance follows from the fact that it allows the classification of the 
physical operators in two sets:\\
\begin{itemize}
\item Dynamical operators, which appear in conjugated pairs, and give a 
central term in their commutators: $[\hbox{ . },\hbox{ . }] \sim 1$.
\item Kinematical operators, which, broadly speaking, do not give this central 
term (the distinction between dynamical and kinematical operators can
be rigorously characterised in terms of a pre-symplectic form constructed from
the group two-cocycle \cite{Aldaya}). Space-time should appear within this 
second set which, in the language of GAQ, is called characteristic subalgebra
and denoted by ${\cal G}_{\Theta}$.
\end{itemize}
\section{The mathematical model}
The first step in the construction of the model consists in selecting an 
appropiate starting group. We are choosing the Virasoro group, in 
abstract terms, for such a group. The reasons for this are, on the one hand, 
that it is a group simple enough in order to be handled and, at the same time, 
rich enough for giving non-trivial answers. On the other hand, Polyakov's 
action for $2D$ gravity \cite{Polyakov} can be derived from it 
(\cite{Alekseev,Miguel}).  

The Virasoro algebra is defined ($h= \frac{c-c'}{24}$) by:
\bea
[ L_n,L_m]=( n-m) L_{n+m}+\frac 1{12}(
cn^3-c'n) \delta _{n,-m}. \nn 
\eea
A formal group law ($l''^m =F^m(l'^n,l^n)$) can be derived from this algebra,
and from this group law the rest of the elements of GAQ (quantization 1-form, 
left- and right-invariant infinitesimal generators...) can be computed. As we 
said before, we are 
particularly interested in the characteristic subalgebra, the kinematical 
degrees 
of freedom. Looking at the commutation relations, we find the two posible 
cases:
\bea
\hbox{i)} \frac{c^{\prime }}{c}\neq r^{2} , r\in Z, \; \Rightarrow \; {\cal G}_\Theta &=& \langle \widetilde{X}_{l^o}^L\rangle \nn \label{r} \\
\hbox{ii)} \frac{c^{\prime }}{c}=r^2, r\in Z, \; \Rightarrow \; 
{\cal G}_\Theta&=&\langle \widetilde{X}_{l^{-r}}^L,\widetilde{X}_{l^o}^L,\widetilde{X}_{l^r}^L\rangle \nn.
\eea
Since we are looking for a space-time with one time dimension and at least
one spatial dimension, the first case is excluded. 
Therefore, our first conclusion is that in order to find a space-time, we must
fall in the second critical case in which ${\cal G}_\Theta$ closes an 
$sl(2,R)$ algebra. In fact, unitarity imposes $r=1$, so that we have
$c=c'$, which is the only case we consider from now on.

The quantum representation is achieved by taking  the set of $U(1)-$complex 
functions
defined on the group as the Hilbert space, and the group acting on it via 
the regular representation. The main problem then is that this representation 
is highly reducible. In principle, this can be solved by taking advantage of 
the trivial commutation among left- ($\widetilde{X}_{l^n}^{L}$) and 
right-invariant ($\widetilde{X}_{l^n}^{R}$) vector fields. This allows us to 
implement the representation with one set of vector fields (for instance, the 
right-invariant ones: $L_n=i(\widetilde{X}_{l^n}^{R})$) and to reduce the 
representation in a consistent way by using the other set, via the 
introduction of polarization equations: $\widetilde{X}_{l^{n\leq 1}}^{L}=0$.

However, in the general case for the Virasoro group, the resulting 
representation is still reducible. We have two solutions to this problem which 
prove to be equivalent. The first one benefits from the existence of a 
vacuum state and consists in taking the orbit of 
the group through this vacuum. The resulting representation is an 
irreducible highest-weight representation, characterised by:
$L_n \mid\! 0\rangle=0, (n\geq -1)$ and $\mid\! \Psi\rangle =L_{n_j}...L_{n_1}
\mid\! 0\rangle, 
(n\leq -2)$. The second solution, which is more natural in the framework of 
GAQ, consists in imposing further polarizations conditions using higher-order
differential operators. This makes pseudo-differential operators come on the 
scene, something which is a source of technical difficulties, however. For the 
sake of simplicity we are not going to consider those cases in which 
higher-order polarizations appear.
The equivalence between the two solutions was proved in \cite{Navarro-Salas}.

With regards to unitarity, the values of $c$ and $c'$ that make unitary the 
representation are \cite{conformal}:
\begin{itemize}
\item
$c \geq 1$, with   $\frac {c-c'}{24} \geq 0$.  
\item
$0<c<1$  with:    
$c= 1- \frac 6{m(m+1)}$ and  
$\frac{c-c'}{24}= \frac {{[(m+1)r - ms]}^2-1}{4m(m+1)}$,
where $1\leq s\leq r \leq m-1$,
and  $m,r,s$  integers with $ m\geq 2$. 
\end{itemize}
But the discrete values of $c$ and $c'$ for $0<c<1$ are precisely those cases 
related with higher-order polarizations and are, therefore, disregarded.

Thus, what we have by now, is a representation of the Virasoro algebra in 
which 
$c=c'$ (in order to find a space-time) and $c>1$ (to have unitarity).
Furthermore, we have classified the operators in two sets: space-time 
operators ($L_{-1}, L_0, L_1$), and dynamical operators ($L_n, \mid n \mid 
\geq 2$), which we shall refer to as {\it gravity} operators in the sequel.
This way, the starting point of the work, the quantum theory, is constructed 
and now we try to accomplish our main objective by identidying a space-time 
structure.

Firstly, we consider the reduction of the Virasoro Hilbert space, 
${\cal H}_{(c,c)}$, under the kinematical $SL(2,R)$ subgroup, obtaining:\\
\bea
{\cal H}_{(c,c)}=\bigoplus_{N} (D^{(N)}-D^{(N-1)})R_{S}^{(N)} \nn 
\eea
where,
\begin{itemize}
\item  $R_{S}^{(N)}$ is a maximal-weight irreducible representation of 
$SL(2,R)$ with Casimir $N(N -$\nolinebreak$1)$. We denote the states in this representation by
$\mid\! N,n\rangle$.
\item $D^{(N)}$ is the dimension of the Virasoro level $N$.
\item The different representations $R_{S}^{(N)}$ are orthogonal. This will 
be important in the physical interpretation because it permits a standard 
quantum mechanical interpretation.
\end{itemize}

Once the representation has been reduced, we proceed to associate a 
space-time with each $SL(2,R)$ representation in the model; that is, as we have
an infinite number of $SL(2,R)$ representations, an infinite 
number of space-times are realised {\it simultaneosly} in our model.

In order to make this association more concrete, we take an specific $SL(2,R)$
representation and construct a $C^*$-algebra by considering all the products 
of the wave functions in the representation. At this point, we can apply a 
theorem by Gelfand and Naimark \cite{Connes}, which allows the reconstruction 
of a manifold from the $C^*$-algebra. The 
problem with this approach is that the mentioned theorem is a rather abstract
tool and the identification of the actual manifold under consideration is a 
difficult task.
Fortunately, we can look at the problem in another way. For this, we consider
again an isolated $SL(2,R)$ representation and notice that we know another 
system with the same Hilbert space but for which the configuration space is 
explicitly known. This system is a particle moving on a two-dimensional AdS 
space-time. Thus, we associate with each $SL(2,R)$ representation a one-sheet
hyperboloid. In fact, this is what one expects to find from a $SL(2,R)$ group,
which is isomorphic to AdS group in two dimensions when imposing the Casimir 
constraint.

Before entering into the physical interpretation, let us make some brief 
comments on the mathematical model we have just introduced. Firstly, in the 
framework of GAQ, it is natural to assign dimensions to the vector fields. In 
our case we fix $[L_n]=(Length)^{-1}$. The commutation relations then 
imply $[c]=Length$ and $[n]=[c']=(Length)^{-1}$. But this represents a problem
when trying to interpret expressions like $c>1$, since we need a scale. To
solve this problem we introduce a constant $a$, such that $[a]=Length$, and 
redefine $n\rightarrow\frac{n}{a}$. Thus, the redefined integers are 
dimensionless.

Another important point is that the Virasoro algebra has a natural notion of 
classical limit which is obtained by making $c\rightarrow\infty$, or in 
redefined terms, $\frac{c}{a}\rightarrow\infty$. In this case, the constant 
$\frac{a}{c}$ behaves as a parameter of a perturbative series, playing the 
role of the Planck constant in our model. This suggests to redefine the 
Virasoro generators: $H_n\equiv \frac{a}{c} L_n$.

The previous introduction of a hyperboloid associated with a given 
representation does not provide a metric structure for space-time. The only 
natural metric we can find inside the model is the one induced from the 
Killing 
metric of $SL(2,R)$, thus providing an AdS space-time. In order to fully 
determine
it, we have to introduce a scale: the {\it radius} of the hyperboloid. It 
is defined from the Casimir in terms of the redefined Virasoro generators:
\bea
\frac{1}{R^2}\equiv {H_0}^2-\frac {1}{2}(H_1H_{-1}+H_{-1}H_1)= \nn \\
=(\frac{a}{c})^2 \frac{N(N-1)}{a^2}= \frac{N(N-1)}{c^2}. \nn \\
\eea
\section{Physical interpretation}
Now we can provide a physical interpretation. As we have stressed, we associate
an AdS$_2$ space-time of radius  $R=\frac{c}{\sqrt{N(N-1)}}$ with each 
$SL(2,R)$ representation.

Given an specific $SL(2,R)$ representation, we interpret each vector in it 
($\mid N,n\rangle$) as a state of the corresponding space-time. Since we are 
dealing with maximal-weight representations, a tower of states of the
space-time is found, where the maximal-weight vector plays the role of 
space-time ground state. $H_0$ is interpreted as the energy and 
$H_{-1}$ ($H_1$) as raising (lowering) space-time operators.
When we consider the dynamical operators, $H_{\mid n\mid\geq 2}$, we notice 
that they do not leave invariant the $SL(2,R)$ representations and therefore
they produce the effect of mixing the different space-times.

The whole picture provided by the model is as follows: we understand by 
{\it Universe} the entire ensemble of different space-times which are realised
at the same {\it time}. A state of the {\it Universe} is a particular state in
the Virasoro representation, that is, an specific linear combination of states
in different $SL(2,R)$ representations or, in other words, a quantum 
superposition of different space-times. The coefficients of the linear 
combination define a weight distribution of space-times.
The question about the radius of the Universe makes no real sense, since we 
have space-times of different radii. The meaningful question is about the 
probability for the Universe to have a certain radius, and then it is 
essential the orthogonality of the $SL(2,R)$ representations, which allows
the definition of orthogonal projectors. Finally, the effect of {\it gravity}
is that of mixing the different space-times, thus changing the weight 
distribution of space-times.

As far as the classical limit is concerned, we can consider the limit
$c\rightarrow\infty$ and find 
that for every space-time the energy of the ground state tends 
to zero ($Energy(\mid\! N,0\rangle)=\;\frac{N}{c}\rightarrow 0$), and the radius
to infinity ($R=\frac{c}{\sqrt{N(N-1)}}\rightarrow\infty$). There is no 
physical way to distinguish between the different space-times in this limit 
and it 
makes sense, accordingly, to identify them. In order to do that, we define the
equivalence relation, $\mid\! N,n\rangle\sim$
\linebreak
$\mid\! N',n\rangle$, and take the 
quotient ${\cal H}\equiv {\cal H}_{(c,c)}/\sim$. The problem with this 
quotient 
is that the dynamical operators are ill-behaved in it; in fact, they are 
multivalued.
To avoid this trouble, we impose to these operators to act trivially on 
the states (this is only consistent with the commutation relations in the 
classical limit, as pointed out by S. Carlip after the talk): 
$H_{\mid n \mid\geq 2}\mid \Psi\rangle =0$.

Hitherto, we have considered the Virasoro group as an abstract one, but
at this point we can look at it as a diffeomorphism group, so that the 
previous constraints are precisely the classical diffeomorphisms constraints.
What we find is that diffeomorphism invariance is only recovered in the 
classical limit, while diffeomorphisms have a dynamical content at the quantum 
level.     

\section{Conclusions}
\begin{itemize}
\item We have constructed a space-time notion from a quantum theory but only 
for the critical value of the Virasoro anomaly ($c=c'$). Out of this value 
space-time makes no sense, even though the quantum theory does exist.
\item The model presents the {\it Universe} as a quantum superposition of 
different space-times which are mixed by {\it gravity} modes.
\item We have implemented a model which exhibits a quantum breakdown of 
diffeomorphism invariance through the appearance of physical (non-gauge) 
degrees of freedom through an anomaly process. General covariance is recovered
only in the classical limit. 
\end{itemize}


\begin{thebibliography}{9}

\bibitem{QG0899} V.Aldaya, J.L.Jaramillo, gr-qc/9907071.

\bibitem{Aldaya} V.Aldaya and J.A. de Azc\'arraga, J. Math. Phys.,  23
                  (1982) 1297;
                 V. Aldaya, A. Ram\'\i rez and J. Navarro-Salas, Commun. Math.
                 Phys.  121 (1989) 541;
                 V. Aldaya, R. Loll and J. Navarro-Salas, Phys.Lett. 
                 B225 (1989) 340.

\bibitem{Polyakov} A.M.Polyakov, Mod. Phys. Lett.  A11 (1987) 893. 

\bibitem{Alekseev} A.Alekseev and S.Shatashvili, Nuc. Phys.  B323 (1989)
                    719.
                   
\bibitem{Miguel}  V.Aldaya, J.Navarro-Salas and M.Navarro, Phys.Lett. 
                  B260 (1991) 311.

\bibitem{conformal} V.G. Kac, Lecture Notes in Physics  94 (1979) 441;
 B.L. Feigin and D.B. Fuchs, Funct. Anal. Appl. 16 (1982) 114;
 C.B. Thorn, Nucl. Phys.  B248 (1984) 551;
 P. Di Francesco, P. Mathieu, D. S\'en\'echal,
 Conformal  field theory, Springer-Verlag, 1997.
 

\bibitem{Navarro-Salas} V.Aldaya, J.Navarro-Salas, Commun. Math. Phys. 
                         126 (1990) 575;
                        V.Aldaya, J.Navarro-Salas, Commun. Math. Phys.
                         139 (1991) 433. 

\bibitem{Connes} Alain Connes,  Noncommutative geometry, Academic Press,
                1994.



\end{thebibliography}
\end{document}